\documentclass[twocolumn]{aastex631}

\shorttitle{Variability of Known Exoplanet Host Stars}
\shortauthors{Emilie R. Simpson et al.}

\begin{document}

\title{Variability of Known Exoplanet Host Stars Observed by TESS}

\author[0000-0003-0447-9867]{Emilie R. Simpson}
\affiliation{Department of Earth and Planetary Sciences, University of California, Riverside, CA 92521, USA}
\affiliation{SETI Institute, Mountain View, CA  94043, USA}
\email{esimp005@ucr.edu}

\author[0000-0002-3551-279X]{Tara Fetherolf}
\altaffiliation{UC Chancellor's Fellow}
\affiliation{Department of Earth and Planetary Sciences, University of California, Riverside, CA 92521, USA}

\author[0000-0002-7084-0529]{Stephen R. Kane}

\author[0000-0002-3827-8417]{Joshua Pepper}
\affiliation{Department of Physics, Lehigh University, 16 Memorial Drive East, Bethlehem, PA 18015, USA}

\author[0000-0003-4603-556X]{Teo Mo\v{c}nik}
\affiliation{Gemini Observatory/NSF's NOIRLab, 670 N. A'ohoku Place, Hilo, HI 96720, USA}

\author[0000-0002-4297-5506]{Paul A. Dalba}
\altaffiliation{Heising-Simons 51 Pegasi b Postdoctoral Fellow}
\affiliation{Department of Astronomy and Astrophysics, University of California, Santa Cruz, CA 95064, USA}
\affiliation{SETI Institute, Mountain View, CA  94043, USA}


\begin{abstract}

Both direct and indirect methods of exoplanet detection rely upon detailed knowledge of the potential host stars. Such stellar characterization allows for accurate extraction of planetary properties, as well as contributing to our overall understanding of exoplanetary system architecture. In this analysis, we examine the photometry of 264 known exoplanet host stars (harboring 337 planetary companions) that were observed during the TESS Prime Mission. We identify periodic signatures in the light curves of these stars and make possible connections to stellar pulsations and their rotation periods, and compare the stellar variability to the published planetary orbital periods. From these comparisons, we quantify the effects of stellar variability on exoplanet detection, confirming that exoplanets detection is biased toward lower variability stars, but larger exoplanets dominate the population of exoplanets around variable stars. Exoplanet detection methods represented among these systems are distinct between stellar spectral types across the main sequence, though notable outliers exist. In addition, biases present in both the sourced data from TESS and the host star selection process, which strongly influences the representation of both stellar and planetary characteristics in the final populations. We also determine whether the host star’s photometric variability affects or mimics the behavior or properties of the system's planets. These results are discussed in the context of how the behavior of the host star is responsible for how we observe exoplanet characteristics, most notably their radii and atmospheric properties, and how the activity may alter our measurements or impact the evolution of planetary properties.

\end{abstract}



\section{Introduction}
\label{intro}

Over the course of several decades, the transit and radial velocity (RV) techniques have been used to discover the vast majority of known exoplanets. With the prolific findings of former and current transiting exoplanet missions such as {\em Kepler} \citep{borucki2010a}, K2 \citep{howell2014}, and the Transiting Exoplanet Survey Satellite \citep[TESS;][]{ricker2015,guerrero2021}, the catalog of candidate and confirmed exoplanets continues to grow. Included in the inventory of stars monitored by TESS are numerous known exoplanet hosts \citep{wong2020e,kane2021b,wong2021,kane2023}, resulting in the discovery of transits for several known RV planets \citep{kane2007b,kane2020c,pepper2020,delrez2021}, as well as new planetary companions in previously established exoplanet systems \citep{huang2018,teske2020}. Since the transit and RV methods of exoplanet detection are indirect, the derived properties of the detected planets rely upon a thorough characterization of the host star \citep{roberts2015c,wittrock2017,jiang2020}. Stellar parameters such as age, composition, size, and effective temperature can affect the formation and stability of a planetary system, as well as the resulting properties of the indirectly detected planets \citep{vanbelle2009a,kane2018a}. As such, characterization of exoplanet host stars remains a critical component of inferring fundamental planetary system properties and their architectures \citep{ford2014,winn2015}.

It has long been known that various types of stellar variability can affect the efficiency and efficacy of exoplanet detection surveys, both in the RV \citep{saar1998,chaplin2019} and photometric \citep{walkowicz2013b,howell2016a} regimes. Some forms of periodic photometric  stellar variability, such as stellar pulsations and rotational star-spot modulation, can have serious consequences for RV and transit exoplanet detection, essentially contributing to the noise floor of the observations \citep{desort2007,cegla2014,andersen2015,korhonen2015}, motivating efforts to disentangle stellar activity from exoplanet signatures \citep{aigrain2012,rajpaul2015,chaplin2019,zhao2022}. Furthermore, photometric variability, such as that caused by stellar activity, can impact the study of exoplanet atmospheres, including data acquired through transmission spectroscopy \citep{zellem2017,cauley2018} or phase variations \citep{serrano2018}. Planetary atmospheric formation, composition, and retention is another prevalent example, since stellar activity can determine whether the exoplanet can retain the atmosphere \citep{ribas2005,roettenbacher2017,kane2020d}, as well as contribute towards a deeper understanding of long-term planetary habitability \citep{lammer2007,segura2010,kopparapu2014,vanderburg2016b}. Consequently, the variability of numerous exoplanet host stars have been studied in detail, including investigations of long-term magnetic cycles \citep{metcalfe2010,kane2011e,dragomir2012a,henry2013,metcalfe2013}. Fortunately, the sky coverage of TESS has enabled a more systematic approach to evaluating the variability of exoplanet hosts, and with greater photometric precision than previous ground-based approaches \citep{ricker2015}.

In this paper, we present results regarding a photometric modulation study of known exoplanet host stars, including their representation in different stellar spectral types and how the variability affects their exoplanets. We include a description of the implications of these results on confirmed and candidate exoplanets, as well as how these findings can be applied to future exoplanet discoveries. In Section~\ref{data}, we describe the data acquisition and detection of stellar variability signatures present in the 2-min cadence data from the TESS Prime Mission. We present the variable stars and their known exoplanets that we include in our sample in Section~\ref{population}. We present the results from a population analysis, with emphasis on both the stellar and planetary properties of the studied systems. Sufficiently variable known host systems are analyzed for both their stellar and planetary characteristics to determine whether any correlations exist due to the variable nature of the host star in Sections~\ref{stellar} and~\ref{planetary}. We provide concluding remarks including suggestions for future work in Section~\ref{conclusions}.


\section{Data Analysis}
\label{data}

\begin{figure*}[t]
    \centering
    \includegraphics[width=0.8\linewidth]{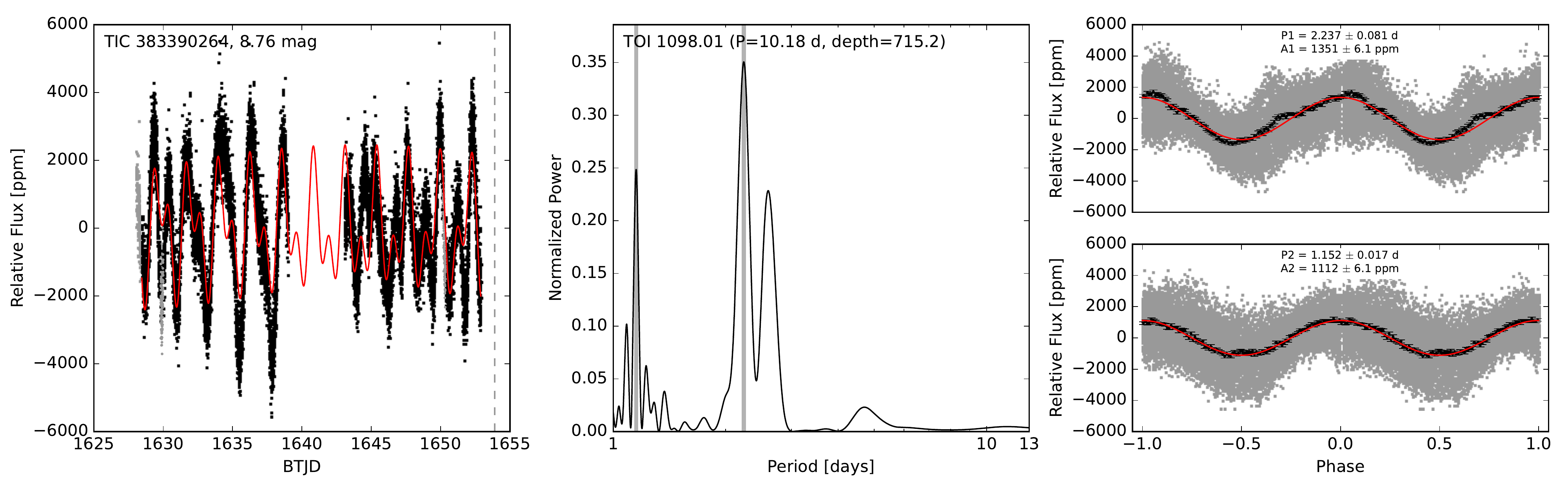}
    \includegraphics[width=0.8\linewidth]{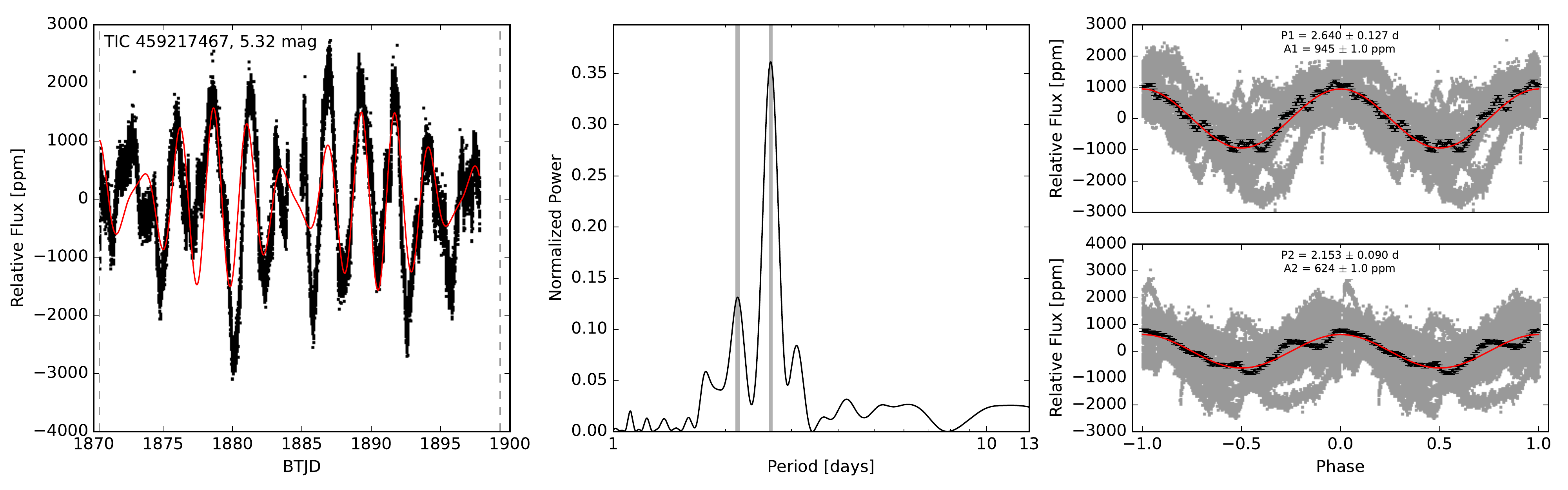}
    \includegraphics[width=0.8\linewidth]{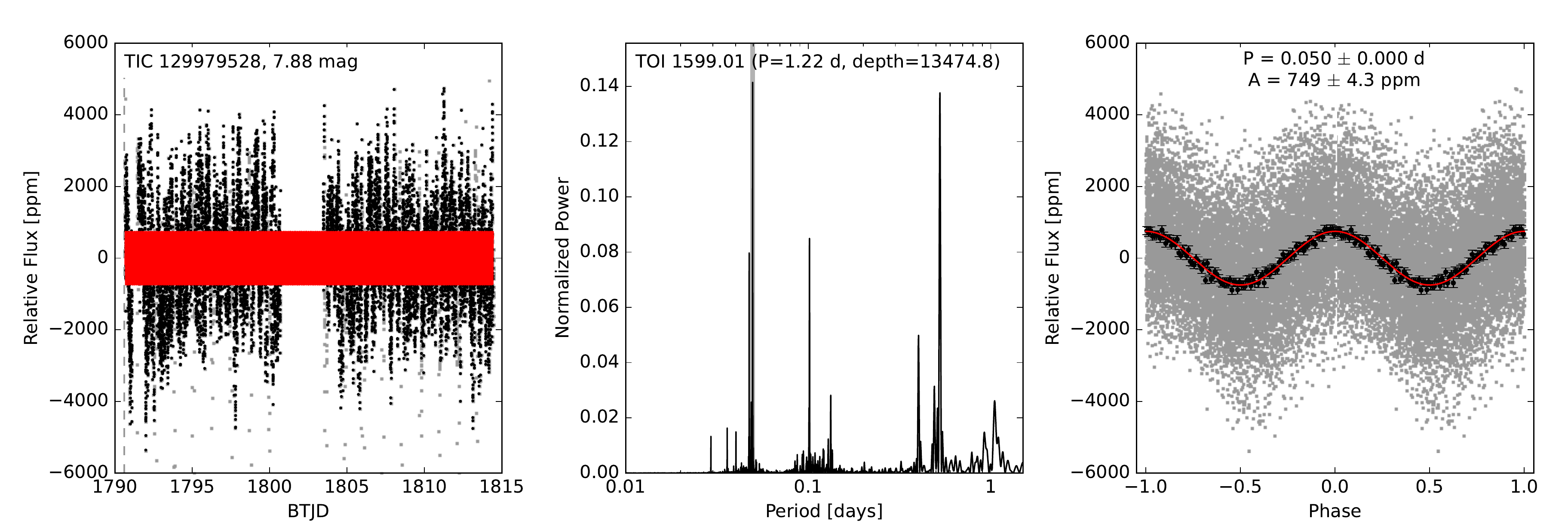}
    \includegraphics[width=0.8\linewidth]{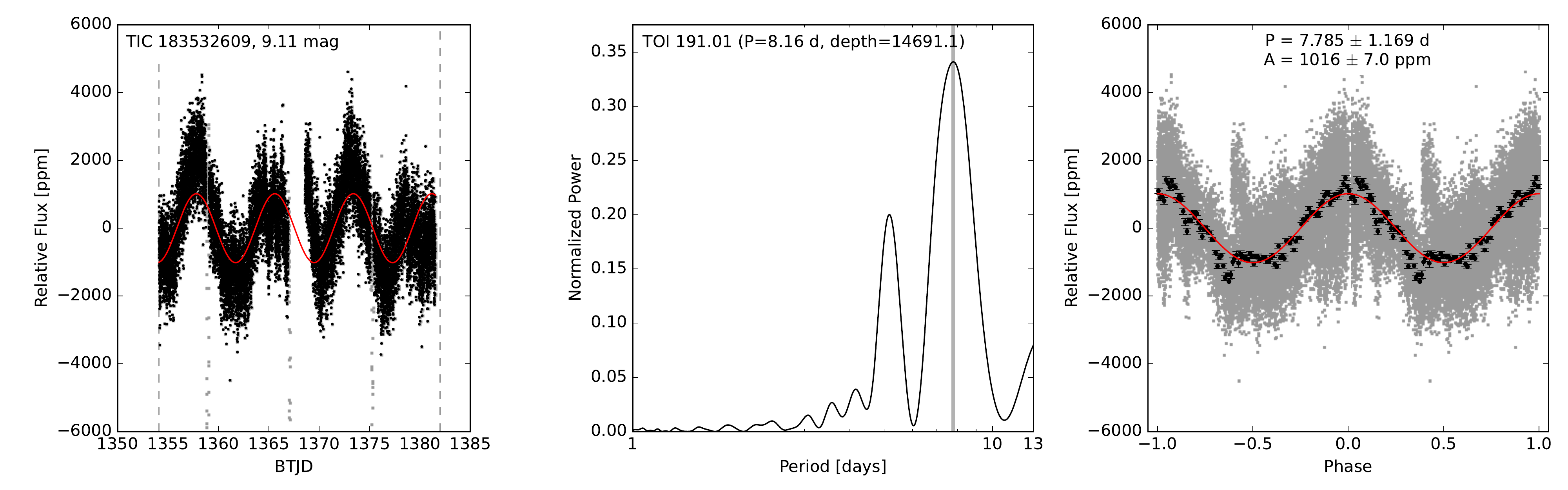}
    \caption{
    Variability analysis results for four known exoplanet hosts (from top to bottom): HD 110082, HD 112640, WASP-33, and WASP-8. Left: the normalized TESS lightcurve with a red sinusoidal fit line, TIC ID, and TESS magnitude. Gray points indicate data that has been removed from the analysis due to being flagged as a transit event or $>$5$\sigma$ outlier. Center: the L-S periodogram with the strongest one or two periodicities indicated by a gray vertical line. When available from the TOI catalog \citep{guerrero2021}, the associated TOI number, orbital period, and transit depth are listed. Right: the phase-folded lightcurve, with a red curve indicating the sinusoidal fit. The gray points show all data included in the analysis, and the black points show the binned data. The best-fit period of variability and flux amplitude are also listed.}
    \label{fig:examples}
\end{figure*}

We utilize data from the TESS Prime Mission, which we briefly describe here but otherwise refer the reader to \citet{ricker2015} for more information. The TESS spacecraft has a 13.7 day orbit that is in a 2:1 resonance with the Moon and observes a $94^\circ\times24^\circ$ segment of the sky with its four optical CCD cameras in 27.4 day increments. During its Prime Mission (2018~July~25--2020~July~04), TESS observed $\sim$70\% of the sky at 30-min cadence and targeted $\sim$230,000 stars at 2-min cadence observations. Photometric measurements and stellar properties are available for each target star in the TESS Input Catalog: \dataset[DOI: 10.17909/fwdt-2x66]{http://dx.doi.org/10.17909/fwdt-2x66} \citep[TICv8;][]{stassun2019}. We utilize the 2-min cadence photometry that are publicly available through the Mikulski Archive for Space Telescopes\footnote{\url{https://archive.stsci.edu/}} (MAST; \dataset[DOI: 10.17909/t9-nmc8-f686]{http://dx.doi.org/10.17909/t9-nmc8-f686}) and were processed by the Science Processing Operations Center (SPOC) pipeline \citep{jenkins2016}, known as the pre-data conditioning simple aperture photometry (PDCSAP) light curves. The PDCSAP light curves are the output from applying co-trending basis vectors to the raw aperture to correct the photometry for common instrument systematics.

Exoplanet host stars are determined to be variable if they are included in the variability catalog developed by \citet{fetherolf2022}. A periodic photometric variability search was performed using the stars observed at 2-min cadence during the TESS Prime Mission (Cycles 1 and 2) that were brighter than $T_\mathrm{mag}=14$ and had less than 20\% flux contamination from neighboring stars. A Lomb-Scargle (L-S) periodogram analysis \citep{lomb1976, scargle1982} was performed to search for periodic variability signals on timescales of 0.01--13\,days, which covered periodic signals up to half of the observing baseline of individual TESS sectors. Based on a visual inspection, variability was determined to be significant when the periodogram peak exceeded 0.01 in normalized L-S power. Variations that could possibly be attributed to spacecraft systematics or those that were limited by the timescale of TESS observations were carefully removed from the catalog, such that the contamination rate of non-variable stars is estimated to be $\sim$21\% at most. From this population of 90,000 variable stars, 264 stars were known to host exoplanets as of November 7th, 2022 \citep{nea}. A Lomb-Scargle (L-S) periodogram analysis \citep{lomb1976, scargle1982} and phase-folded lightcurve analysis provided by \citet{fetherolf2022} is then implemented through the TESS Stellar Variability Catalog (TESS-SVC) High-Level Science Product
(DOI: \dataset[10.17909/f8pz-vj63]{\doi{10.17909/f8pz-vj63}}). This resulted in 337 confirmed exoplanet targets for the overall variability analysis of known hosts.

Figure~\ref{fig:examples} shows examples of four known exoplanet hosts from our sample that exhibit detectable variability signals. Sinusoidal fits to the unfolded and phase-folded lightcurves, shown in red in the left and right panels, verify that the variability is periodic in nature, as opposed to being the result of a flare or other single-event occurrence. The center figure is a Lomb-Scargle periodogram with vertical grey lines indicating the one or two periodicities that were identified to have the highest power. Two periodicities are selected when both are significant to at least 0.1 normalized LS power and there is at least a 25\% improvement in a double-sinusoidal model fit to the full light curve. Below we provide a brief description for each of the example systems shown in Figure~\ref{fig:examples}. These systems were selected for the variety of variability behaviors they exhibit across the scope of the variability catalog.

HD~110082 (TIC~383390264) is a young ($0.250^{+0.050}_{-0.07} Gyr$) oscillating F-type star, and home to a Neptune-sized planet discovered via TESS photometry \citep{tofflemire2021}. HD~110082 has a reported stellar rotational period of $2.34\pm0.07$ days, which was calculated from HD~110082's flare-masked light curve and a scaleable Gaussian process. This compares well with one of the two calculated stellar variability periods of $2.237\pm0.081$ days. This system had been previously considered as a member of the Octans Association \citep{murphy2015}, a young stellar moving group that is valuable for studying circumstellar disk and planetary evolution.

HD~112640 (TIC~459217467) is a K-type pulsating star that harbors a giant planet with a minimum mass of 5~$M_J$ in a $\sim$613~day orbit \citep{lee2020}. The discovery via the RV method required additional analysis steps to ensure that the variations were not caused by stellar activity. To account for long-period, low-amplitude RV variations caused by surface activity or stellar rotation, the discovery team used a line-broadening model \citep{takeda2008} before implementing their Lomb-Scargle periodogram analysis. Their resulting upper-limit RV rotational period is 545 days. The orbital period of the planet is substantially larger (at $613.2\pm5.8$ days) than our detected variability periods of 2.640 and 2.153 days. It can then be assumed that the periodic photoemtric signals have little to no correlation with the detected RV planetary signatures..

WASP-33 (TIC~129979528) is a system with a bright A-type main sequence host star and an inflated giant planet in a $\sim$1.21~day orbit. It is also the only system currently known with a hot Jupiter orbiting a $\delta$ Scuti star. This system is known for the difficulties in data analysis, since the stellar pulsations interfere with the extraction of planetary signatures via both transit and RV measurements \citep{colliercameron2010, vonessen2014, turner2016, zhang2018a}. A study done by \citep{vonessen2014} revealed eight separate pulsation periods, with the dominant pulsation period of 0.05 days lining up with our reported variability period.

WASP-8 (TIC~183532609) is a system that is home to two planets, WASP-8b and WASP-8c. WASP-8b has a mass of $2.54\pm0.33$~$M_J$ and an orbital period of $8.158720\pm{0.000015}$ days while WASP-8c has a mass of $9.45^{+2.26}_{-1.04}$~$M_J$ and an orbital period of $4323^{+740}_{-380}$ days. The host is a bright ($V=9.8$) G-type star with a reported $v \sin i$ measurement of $2.0\pm0.6$~km/s. This system was discovered by \citet{queloz2010}, and is especially notable for the star's high proper motion and planetary orbit misalignment relative to the spin-axis of the host star \citep{bourrier2017b}, resulting in a retrograde orbit of the planet. Our calculated photometric variability period for WASP-8 is $7.785\pm1.169$~days, the cause of which may be related to rotational modulations, but is difficult to confirm from the $v \sin i$ value due to the misalignment of the planetary orbit.


\section{Sample Overview}
\label{population}

We will now focus on stellar parameters that may directly influence the stellar flux received by the planet and subsequent planetary evolution. These stellar properties include the effective temperature, stellar radius, variability period, and variability amplitude. The stellar parameters used in our analysis were sourced from the TICv8 catalog, and are current as of 2022 November 7 \citep{nea}. Luminosity values were independently calculated using the TICv8 stellar radii and effective temperatures.


\begin{deluxetable*}{ccccccc}
\tablecaption{Known Hosts Table}
\tablewidth{700pt}
\tabletypesize{\scriptsize}
\tablehead{
\colhead{TIC ID} &
\colhead{Planet Name} &
\colhead{$P_\mathrm{orb}$} &
\colhead{$P_\mathrm{var}$} &
\colhead{$T_\mathrm{eff}$} &
\colhead{$\log g$} &
\colhead{$V_\mathrm{mag}$} \\ [-0.3cm]
\colhead{} &
\colhead{} &
\colhead{(days)} &
\colhead{(days)} &
\colhead{(K)} &
\colhead{} &
\colhead{}
}
\startdata
396697266 & HD 27894 b & $847.0\pm20.0$ & $5.215\pm0.373$ & $4912\pm111$ & $4.486\pm0.082$ & $9.360\pm0.03$ \\
396697266 & HD 27894 c & $1.36002854\pm0.00000062$ & $5.215\pm0.373$ & $4912\pm111$ & $4.486\pm0.082$ & $9.360\pm0.03$ \\
396697266 & HD 27894 d & $452.8^{+2.1}_{-4.5}$ & $5.215\pm0.373$ & $4912\pm111$ & $4.486\pm0.082$ & $9.360\pm0.03$ \\
12723961 & HD 212771 b & $883.0^{+32.4}_{-13.8}$ & $0.6692\pm0.007$ & $5003\pm25$ & -- & $7.600\pm0.03$ \\
180695581 & TOI-1807 b & -- & $0.6319\pm0.007$ & $4612\pm99.7$ & $4.562\pm0.084$ & $10.000\pm0.03$ \\
116242971 & KELT-12 b & $158.991\pm1.44$ & $1.072\pm0.016$ & $6350\pm134$ & $3.868\pm0.083$ & $10.585\pm0.006$ \\
270501383 & HD 205739 b & -- & $11.97\pm2.786$ & $6308\pm108$ & $4.258\pm0.079$ & $8.560\pm0.03$ \\
612908 & HD 30856 b & $380.7\pm1.4$ & $1.46\pm0.033$ & $4895\pm25$ & -- & $7.910\pm0.03$ \\
138764379 & HD 94834 b & $2.423804\pm0.000008$ & $0.05879\pm0.0001$ & $4798\pm25$ & -- & $7.600\pm0.03$ \\
456905162 & HD 10697 b & $1.4811235\pm0.0000011$ & $1.903\pm0.052$ & $5600\pm103$ & $3.905\pm0.069$ & $6.283\pm0.023$ \\
\enddata
\tablecomments{Listed are the first ten objects in our catalog. The complete table will be available with the online manuscript.}
\end{deluxetable*}

\begin{figure}
    \includegraphics[width=\linewidth]{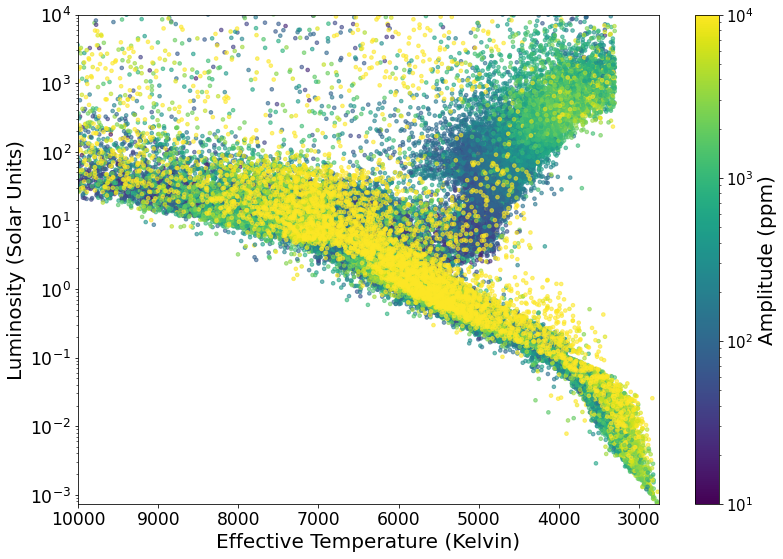}
    \includegraphics[width=\linewidth]{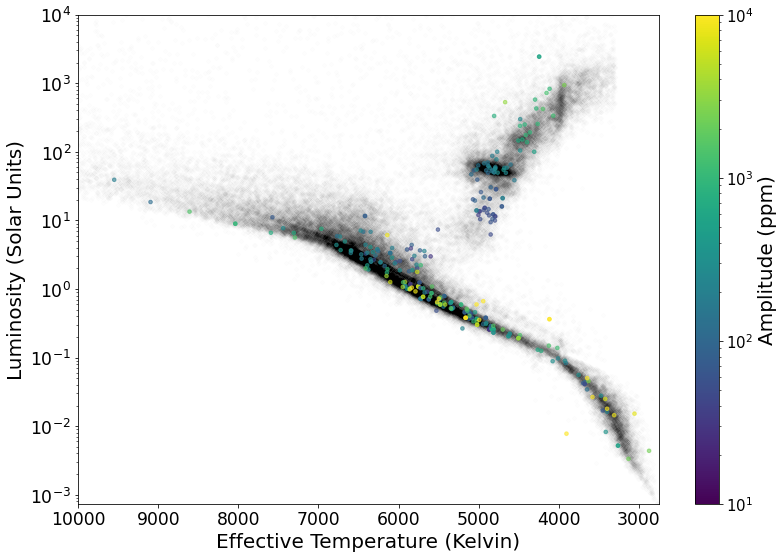}
    \caption{Top: The initial population of variable stars \citep{fetherolf2022} colored by their measured photometric variability amplitude in parts per million (ppm). Bottom: The final population of variable known exoplanet host stars.}
    \label{fig:star_pop}
\end{figure}

Figure~\ref{fig:star_pop} shows our sample of stars on the HR diagram, the top panel displays the initial all-sky variable star population \citep{fetherolf2022} and the bottom panel displays the final variable known hosts population for this work. The graphs are colored by the measured photometric amplitude in parts per million.

\begin{figure}
    \includegraphics[width=\linewidth]{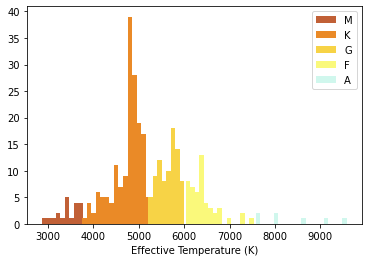}
    \includegraphics[width=\linewidth]{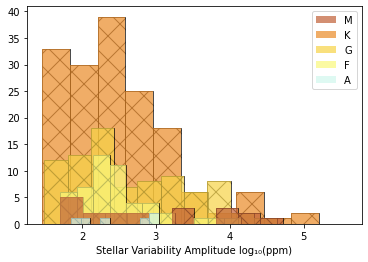}
    \caption{Top: A histogram of the known host population's effective temperature, separated by spectral type. O- and B-type stars are not present within this sample, so they are not represented within this graph. Bottom: A histogram of the known host population's stellar variability amplitude in $log_{10}$, separated by spectral type.}
    \label{fig:histograms}
\end{figure}

Figure~\ref{fig:histograms} shows our sample of stars represented in a set of histograms. The top histogram shows the distribution of effective temperatures, which are further separated by spectral type. The bottom histogram shows the distribution of stellar variability amplitudes, also separated by spectral types. The stellar variability amplitude histogram uses logarithmic bin sizes to improve population visibility.

\citet{fetherolf2022} found that many of the high-amplitude variability targets tend to be eclipsing binary systems, which is consistent with eclipsing binaries tending to sit at higher luminosities than single stars on the main sequence \citep[e.g.,][]{babusiaux2018}. Our sample of known exoplanet host stars mostly consists of variable stars with medium- to low-amplitude variability, which may be related to the selection bias against variable stars when searching for exoplanets. The photometric variability of the exoplanet host stars also tends to be smoothly sinusoidal in nature, which would be consistent with rotational modulations or stellar pulsations. There is also a gap in the main sequence for our sample where the K-dwarf stars (3700K--5200K) should be. While K-dwarfs are overall abundant in the observable sky and the TESS FFIs, there exists a selection bias within TESS target selection for the 2-min cadence observations that favors G-dwarfs and M-dwarfs over K-dwarfs \citep{kaltenegger2009, ciardi2015} to promote exoplanets within the habitable zone of their host star.

To gain a better understanding of the variability of the known host star population, we conducted separate analyses of the variable host stars and their exoplanets. These two population studies can reveal a more complete profile of these variable stars and how they affect their exoplanets.


\section{Stellar Properties}
\label{stellar}

All stellar data within this study was sourced from the TICv8 catalog. These values, paired with the more extensive photometric and spectroscopic data available from the NASA Exoplanet Archive supplement the overall variability analysis. Upon first glance, the variable host stars more or less lie within the expected range of stellar spectral types, with most targets being along the main sequence, but some populate the giant branch. The red giant branch is heavily populated by variable stars. Some targets populate the very bottom of the sub-giant branch. O- and B-type stars ($>$10,000\,K) are rare, and therefore not well-represented in any former or current photometric mission data. Note that there are numerous minor differences (as low as within 0.0134\%) between the stellar values extracted from the TICv8 catalog and those reported by the NASA Exoplanet Archive. However, some of these stellar property values can have average differences as high as 12--46\%. The discrepancies may be the result of applying blackbody models to giant stars, leading to an increase in discrepancies as the values stray away from the main sequence \citep{stassun2019}.

We will now discuss the discrepancies in reported stellar parameters between the TICv8 catalog and the NEA. Among the 264 systems, 262 host stars have stellar parameters present in both the NEA and the TICv8 catalog. 192 of these stars are considered within the main sequence, while the remaining 70 are considered evolved stars. For the main sequence stars, discrepancies in effective temperature range between ~0.001\% and ~10.746\% of one another. The majority of these main sequence stars have luminosity discrepancies (which are calculated via their reported stellar radius values) ranging between ~0.027\% and ~41.660\%. Two main sequence targets that exceed this value are TOI-1227 (~64.660\%) and XO-6 (~84.889\%). Discrepancies in effective temperature for evolved stars range between ~0.059\% and ~8.715\%, with one target(HD 95127) having a ~12.253\% discrepancy. Luminosity discrepancies range between ~0.090\% and ~48.402\%, with one target (BD+20 274)having an abnormally large discrepancy at ~84.118\%.

\begin{figure*}
    \includegraphics[width=\textwidth]{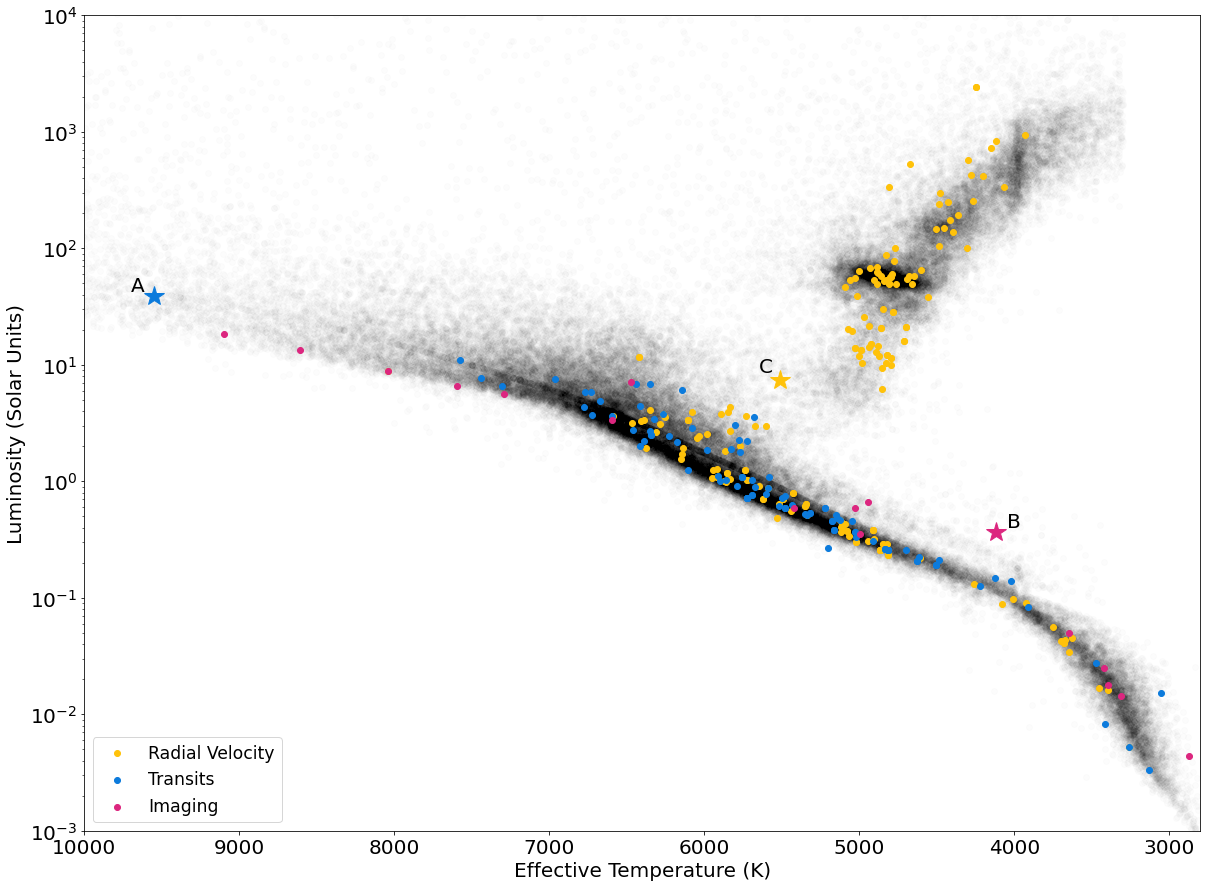}
    \caption{The variable stars with known exoplanets colored by the detection method of their planets. The starred points are the locations of the unique cases KELT-9 (A), PDS~70 (B), and HD~106270 (C).}
    \label{fig:detection}
\end{figure*}

Figure~\ref{fig:detection} displays a HR diagram of the full sample, where the known exoplanet hosts are color-coded by their detection method. The detection method associated to a target can then imply characteristics of the host star or the planet, such as planetary orbit inclination, stellar brightness, and mass ratio. We can see here how observational biases favor certain stellar spectral types. Planets around giant stars are more likely to be found via their gravitational influence (i.e., RV method) rather than transit photometry. High-mass stars will have an abundance of imaging planets, likely due to their young ages that make the planets easier to observe in the infrared. Low-mass stars are especially subject to influence from gravitational pull from their planets. Transit photometry is also effective for these stars due to the radius ratio; i.e., small planets are more easily found around small stars. The relative lack of K dwarf host stars may be due to the sample selection bias of exoplanet surveys that tend to be optimized toward either G dwarfs or M dwarfs \citep{kaltenegger2009,ciardi2015}.

There are outliers marked by starred points within this plot that are unusual for what one expects these detection methods to favor. These targets may lend themselves to a more diverse understanding of the star-planet connections that exist in this population. These are discussed in the following subsections.


\subsection{KELT-9}

KELT-9 is the hottest known host star within this sample, located at the hot end of the main sequence. KELT-9b was initially detected via transit photometry, but was then confirmed with the RV method by the discovery team \citep{gaudi2017}. The exoplanet is an ultra-hot Jupiter that also exhibits a significant atmospheric phase curve \citep{wong2020d}. The rotational velocity $v\sin i$ measured by the discovery team was $111.4\pm1.3$ km/s, which is consistent with our measured photometric variability period of $0.494 \pm 0.003$ days when assuming the stellar rotation is aligned with our line-of-sight and using the reported stellar radius of 0.23 $R_\odot$.


\subsection{PDS~70}
PDS~70 is a K7-type, $5.4 M_{yr}$ star primarily known for its young circumstellar disk. It is represented within Figure~\ref{fig:detection} as a starred imaging target. It is a host to two planets: PDS~70b \citep{keppler2018} and PDS~70c \citep{haffert2019}. These planetary companions were found via point-source detection within near-infrared images of the protoplanetary disk. PDS~70b has an approximate orbital period of 43,500 days. PDS~70c has no reported orbital period. The reported variability period for PDS~70 is significant and periodic in nature, so we can safely assume that the protoplanetary disk does not interfere with the quality of the variability measurements.


\subsection{HD~106270}

HD~106270 is a G-type subgiant star that has one planetary companion, HD~106270b \citep{johnson2011}. This Jovian planet was found among seventeen others resulting from the California Planet Search (CPS) focusing on evolved stellar targets from the Hipparcos catalog \citep{leeuwen2007}. HD~106270b has a reported orbital period of  $1888/pm16$ days. There was limited phase coverage for this target due to the duration of the orbital period, which exceeded the duration of observations for the entire study. The calculated variability period for HD~106270 is $6.92\pm0.74$ days. Follow-up observations of this target have further refined the stellar properties, but the properties of the planetary companion require more extensive observation.


\section{Planetary Properties}
\label{planetary}

We now shift our focus to analyze and understand how exoplanets of variable host stars compare against exoplanets with quieter host stars. To begin, we compare our population of variable host star planets against the current population of confirmed exoplanets in a mass-radius diagram and a orbital period-radius diagram.
\begin{figure*}
    \includegraphics[width=.45\textwidth]{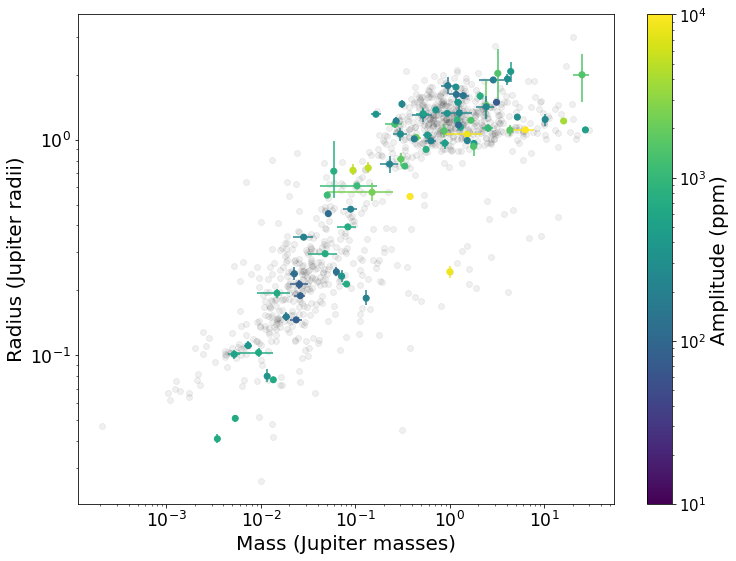}
    \hspace{1cm}
    \includegraphics[width=.45\textwidth]{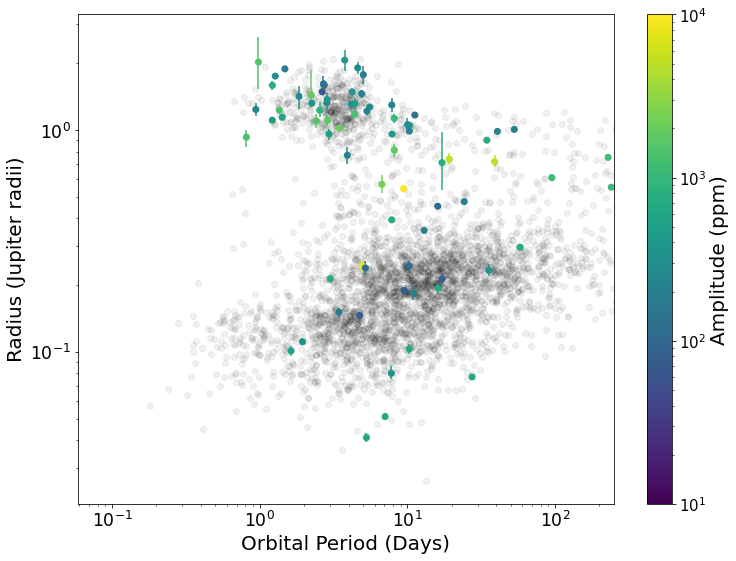}
    \centering
    \caption{Left: A scatter plot that represents the mass-radius relationship of known planets around variable stars. The points are colored by their photometric variability amplitude in ppm. Points in grey represent the overall population of confirmed exoplanets from the NASA Exoplanet Archive as of 20 May 2022. Right: A scatter plot of exoplanet radii and their orbital periods for exoplanets around variable host stars (colored) and the overall population of confirmed exoplanets (grey).}
    \label{fig:relations}
\end{figure*}

Figure~\ref{fig:relations} contextualizes where planets around variable stars lie among the broader population of confirmed exoplanets. The left plot shows their locations on a mass-radius diagram, where the right plot compares radius and orbital period, with the colored points showing the amplitude of the measured stellar variability. There are some features of both the mass-radius plot and the radius-period plot that discern these exoplanets from the entirety of the confirmed exoplanets population. A slight gap can be seen at the location of the Fulton gap \citep{fulton2017} within both plots, which is an observed scarcity of exoplanets between 1.5 and 2 Earth radii, possibly due to photoevaporation-driven mass loss \citep{owen2013a,owen2017c}. Targets with the highest variability amplitude tend to lie within or next to this gap. In addition, there are more higher-amplitude host stars that have higher-radius exoplanets. This is expected due to larger planets being more easily discerned from stars with strong variability. The transit photometry of larger planets against variable stars is easily discerned compared to a smaller target who's transits could be masked by the activity of the star.

These planets are well-distributed against the population of confirmed exoplanets. Cutoffs that exist within the data can be attributed to detection biases; it is of no surprise that a prominent group of high-radius exoplanets dominates our sample of planets around variable stars as they are the easiest to detect against even highly variable stars.


\subsection{A Discussion of False Positives}

The detection of false positives is a practice that is constantly  developed and refined. Identifying and/or remedying false positives in astronomical data has been extensively documented \citep{brown2003,charbonneau2004,torres2004,odonovan2006,latham2009,evans2010,kane2016a,sullivan2017,collins2018}. Finding false positives and removing them from further studies results in more accurate exoplanet population studies and exoplanet occurrence rates. In addition, these false positives reveal unique features in photometry and spectroscopy data that can be flagged as suspicious, which improves the methods of confirming exoplanets.

In this study, we address the possibility of false positives induced by the variability of the host star. These variable stars have a periodic fluctuation within their overall variability that can falsely represent an exoplanet. The false-positive results can be remedied with follow-up analysis, and their discovery has led to improved exoplanet confirmation methods that endure through new developments of planet hunting.

\begin{figure}[h]
\includegraphics[width=\linewidth]{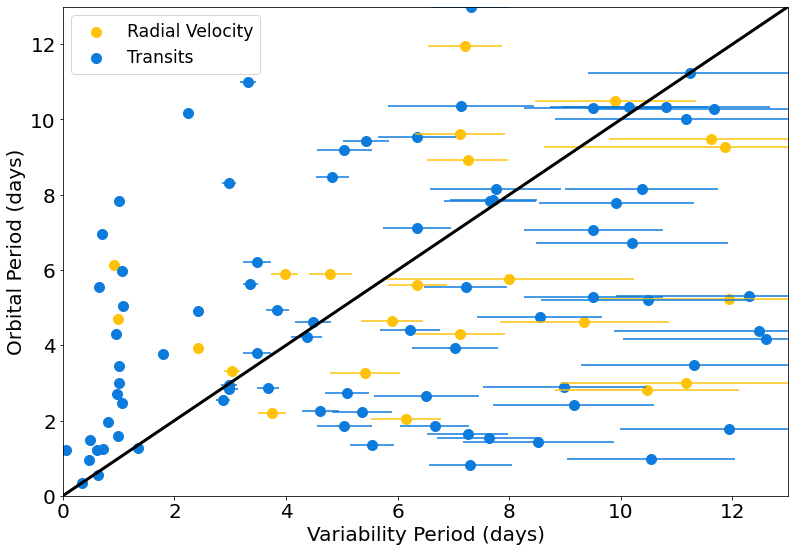}
\caption{A graph comparing the variability period of the host star to the reported orbital period of the planet. The line represents a one-to-one match. The closer a point lies to this line, the closer the variability period and the orbital period match in value. Note that orbital period uncertainties are included in the plot, but are too small to be visible on this scale.}
\label{fig:period-period}
\end{figure}

Figure~\ref{fig:period-period} shows a comparison between a target's reported orbital period and the host star's variability period. Only the targets with orbital periods that are less than 13 days are shown due to the search limit of the variability analysis. Note their proximity to the one-to-one line: the closer a point is to this line, the closer their reported orbital period and the stellar variability period match. There are also a few systems that tend to collect at a variability period of $\sim$1 day, but they have a broad range of exoplanet orbital periods. This is a known artifact of the LS periodogram search performed by \citet{fetherolf2022}, which tends to occur when the detected periodic variability is low in normalized power ($<$0.05).

Out of all the targets represented within this figure, 2.0\% of the reported stellar variability and orbital periods lie within 5\% of each other. Four transiting targets (TOI-677~b, WASP-23~b, HAT-P-17~b, and WASP-105~b) have stellar variability periods and orbital periods that lie within 2\% of one another. The dispersion of these points grows as the stellar variability period increases. This is due to less data being available to verify longer period targets. The uncertainties for the stellar variability period are determined from the width of a Gaussian fit to the LS periodogram peaks. As the variability periods get longer, the peaks in these LS periodograms become broader. These broad peaks then result in higher calculated uncertainty values. Considering that this population represents a small portion of all confirmed exoplanets, it must be considered whether or not this value lies within expectations. In order to determine the probability of these values, we created weighted randomized samples that reflect the distribution of points we see in Figure~\ref{fig:period-period}. The probability distribution for the stellar variability and orbital periods were drawn from an interpolation of their histograms based on those measured in our sample. The probability distributions were then used to produce new randomized datasets of associated orbital periods and stellar variability periods. The chance similarity (within 5\% difference) between the randomized orbital period and stellar variability period was measured for each randomized sample. After 10,000 randomized samples, the average percentage of targets that had a chance similarity between their orbital period and stellar variability period within 5\% was 3.283\%. The randomized chance (3.283\%) of a target's orbital period matching it's stellar variability period exceeds what is observed in the real sample (2.0\%), so we can determine these odds.

Targets with stellar variability periods close in value to their exoplanet's reported orbital period can be considered as false-positive candidates or could be planets that exhibit atmospheric phase variations. Atmospheric phase curves have a characteristic shape and phase that are directly related to the planet's orbit and physical properties \citep[for a recent review, see][]{shporer2017a}. In the case of BD-06 1339b \citep{simpson2022}, the amplitude of the phase curve signal was much larger than what was anticipated from the mass and radius of the planet based on the RV data---thus deeming it a false positive exoplanet. The known exoplanet variable host stars presented in this work went through a visual inspection process of their full and phase-folded light curves to search for atmospheric phase curve candidates and potential false positives. Targets with similar period values were then reviewed to ensure the variability present within the data was astrophysical in nature and not systematic. Some examples of filtering include instances of momentum dumps being confused for periodic variability and leaving out targets with polluted/noisy periodogram results. In the case of BD-06 1339b, besides a close stellar variability period and orbital period match, the phase curve of the normalized flux and the phase curve of the RV signals were close in phase and the flux amplitude was too large compared to the reported planet size.

A primary method for ruling out false positives in the case of a close period match is if the planet is transiting. Figure~\ref{fig:period-period} shows the targets color-coded by their detection method, which reveals 57 more transiting targets with similar orbital and stellar variability periods than RV targets.


\section{Conclusions}
\label{conclusions}

The variability of exoplanet host stars remains a major challenge for the detection of the planets in those systems and can, in rare cases, result in spurious exoplanet signals. Moreover, the resulting change in incident flux at the top of exoplanet atmospheres can influence the climate state of terrestrial planets, depending on the frequency and amplitude of those changes. Consequently, the study of stellar variability is an important topic in the study of both exoplanet detection and characterization. Fortunately, there now exists a large inventory of precision photometry for most of the stars that have known exoplanets, allowing an investigation of the stellar variability and the relationship with other star and planet properties.

The study presented here has made specific use of the TESS Prime Mission photometry to conduct a systematic search for stellar variability. The original population of variable stars, as detected by \citet{fetherolf2022}, was subjected to filtering processes to promote targets with significant periodic variability. We then analyzed a subset population of 264 variable known host stars, along with their 337 associated planetary companions, and searched for correlations between planetary properties and the variability of their host star. We discussed the variable host stars in terms of their stellar parameters, the strength of their variability in conjunction with their status as exoplanet host stars, and the sources of their variability. The planets of these variable stars were discussed in terms of how the variability of their host stars affected their discovery and the overall demographics of the exoplanet population. By utilizing photometry from the TESS spacecraft, we see that variable known hosts populate a range of spectral types, but vary in strength and variability classification in certain sub-regions of the HR diagram.

Overall, we find that the resulting variable known host population is heavily influenced by selection bias, due to relative sensitivity of exoplanet detection method when applied to stars with medium to low-amplitude variability versus those with high-amplitude variability. While we can infer a good understanding of exoplanet demographics among stars with lower-amplitude variability, specific characteristics for exoplanets around stars with high-amplitude variability remain less accurate. This relative lack of information for the properties of exoplanets orbiting stars with high-amplitude variability directly impacts our understanding of how a star's variability may affect exoplanet atmospheres. Certain stellar spectral types favor specific exoplanet detection types, but notable outliers exist and define interesting cases of diversity among star-planet relationships. Exceptions tend to feature variable known exoplanet host stars that have evolved off the main sequence. Our population of variable known hosts falls within the expected probability of having planets whose orbital periods fall close to the star’s variability period in value, creating instances of possible false positive cases that warrant further investigation. While false positives are not the primary focus in the scope of this paper, they are important to discuss in the perspective of how we perceive stellar variability and how it may be misinterpreted as a planetary signal.

The TESS mission is an ongoing effort to detect new planetary systems and additional planets in known systems, extending the baseline of observations to increase sensitivity to longer period orbits. A by-product of these extended observations is the improved capability to detect longer period stellar variability that can affect exoplanet detection and characterization on longer time scales. In general, studies such as the one presented here have the potential to provide a foundation of understanding that can be applied to both current variable planetary systems and those found in future exoplanet missions.


\section*{Acknowledgements}

The authors acknowledge support from NASA grant 80NSSC18K0544, funded through the Exoplanet Research Program (XRP). T.F. acknowledges support from the University of California President's Postdoctoral Fellowship Program. P.D. acknowledges support from a 51 Pegasi b Postdoctoral Fellowship from the Heising-Simons Foundation and a National Science Foundation (NSF) Astronomy and Astrophysics Postdoctoral Fellowship under award AST-1903811. This research has made use of the NASA Exoplanet Archive, which is operated by the California Institute of Technology, under contract with the National Aeronautics and Space Administration under the Exoplanet Exploration Program. This paper includes data collected with the TESS mission, obtained from the MAST data archive at the Space Telescope Science Institute (STScI): \dataset[DOI: 10.17909/t9-nmc8-f686]{http://dx.doi.org/10.17909/t9-nmc8-f686}. Funding for the TESS mission is provided by the NASA Explorer Program. STScI is operated by the Association of Universities for Research in Astronomy, Inc., under NASA contract NAS 5–26555. All of the data presented in this paper were obtained from the Mikulski Archive for Space Telescopes (MAST). STScI is operated by the Association of Universities for Research in Astronomy, Inc., under NASA contract NAS5-26555. Support for MAST for non-HST data is provided by the NASA Office of Space Science via grant NNX13AC07G and by other grants and contracts. This research made use of Lightkurve, a Python package for {\em Kepler} and TESS data analysis \citep{Lightkurve2018}.


\facilities{TESS}

\software{Astropy \citep{Astropy_Collaboration13, Astropy_Collaboration18},
          Lightkurve \citep{Lightkurve_Collaboration18},
          Matplotlib \citep{Hunter07},
          NumPy \citep{Harris20},
          SciPy \citep{Virtanen20}
          }




\end{document}